\documentclass[showpacs,fleqn,nobibnotes]{revtex4}

\usepackage{amsmath}
\usepackage{graphicx}

\def\lsim{\raise0.3ex\hbox{$<$\kern-0.75em\raise-1.1ex\hbox{$\sim$}}}
\def\gsim{\raise0.3ex\hbox{$>$\kern-0.75em\raise-1.1ex\hbox{$\sim$}}}

\def\pom{{I\!\!P}}

\newcommand{\rr}{\mbox{\boldmath $r$}}

\newcommand{\rb}{\mbox{\boldmath $b$}}

\newcommand{\rd}{\mbox{\boldmath $\Delta$}}

\begin{document}

\title{Exclusive $\rho$ and $J/\Psi$ photoproduction in ultraperipheral $pA$ collisions: Predictions of the gluon saturation models for the momentum transfer distributions}
\author{V. P. Gon\c{c}alves $^{1}$, F. S. Navarra $^{2}$  and D. Spiering $^{2}$}

\affiliation{$^{1}$ Instituto de F\'{\i}sica e Matem\'atica,  Universidade
Federal de Pelotas, %\\
Caixa Postal 354, CEP 96010-900, Pelotas, RS, Brazil}
\affiliation{$^{2}$ Instituto de F\'{\i}sica, Universidade de S\~{a}o Paulo, CEP 05315-970 S\~{a}o Paulo, SP, Brazil.}

\begin{abstract}
In this letter we complement previous studies on  exclusive vector meson photoproduction in hadronic collisions presenting a  comprehensive analysis of the $t$ - spectrum measured in exclusive $\rho$ and $J/\Psi$ photoproduction in $pA$ collisions at the LHC. We compute the  differential cross sections considering two phenomenological models for the   gluon saturation  effects and present predictions for $pPb$ and $pCa$ collisions. Moreover, we compare our predictions with the recent preliminary CMS data for the exclusive $\rho$ photoproduction. We demonstrate that the gluon saturation models are able to describe the CMS data at small - $t$. On the other hand, the models  underestimate the few data point at large -- $t$. Our results indicate that  future measurements of the large -- $t$ region can be useful to probe the presence or absence of a dip in the $t$ -- spectrum and discriminate between the different approaches to  the gluon saturation effects.
\end{abstract}

\maketitle

During the last years the study of photon -- induced interactions at hadronic colliders has been strongly motivated by the possibility of constraining the dynamics of the strong interactions at large energies (For a recent review see Ref. \cite{review_forward}).
One of most promising observables is the exclusive vector meson photoproduction cross section \cite{vicbert,vicmag}, which is driven by the gluon content of the target (proton or nucleus) and is strongly sensitive to non-linear effects (parton saturation).
Such expectation has motivated the analysis of  exclusive $\rho$, $\phi$, $J/\Psi$, $\Psi(2S)$ and $\Upsilon$ photoproduction in $pp$, $pA$ and $AA$ collisions at RHIC and LHC energies considering different theoretical approaches for the treatment of the QCD dynamics and for the vector meson wave function (See, e.g., Refs.   \cite{vicmag_varios,bruno,schafer,guzey,jones,run2,armesto,contreras}). In particular, the recent study performed in Ref. \cite{run2} indicated that a global analysis of the experimental data for the rapidity distributions of all these different final states will be necessary to discriminate between the distinct theoretical approaches. On the other hand, the results presented in Refs. \cite{armesto,nos_tdist} indicate that the study of the squared  momentum transfer ($t$) distributions is an important alternative to probe the QCD dynamics at high energies. These distributions are expected to provide  information about the spatial distribution of the gluons in the hadron and about fluctuations  of the  color fields (See e.g. Ref. \cite{heike}). In Ref. \cite{nos_tdist} we have presented predictions for the $t$ - spectrum measured in the exclusive vector meson photoproduction considering $pp$ and $PbPb$ collisions at the LHC. Our goal is this letter is twofold. First, to complement that study and present, for the first time, predictions for the momentum transfer distributions measured in  exclusive $\rho$ and $J/\Psi$ photoproduction in $pPb$ collisions considering two phenomenological models for the treatment of the gluon saturation effects. Second, to present a comparison of gluon saturation predictions with the recent (preliminary) CMS data on  exclusive $\rho$ photoproduction in ultraperipheral $pPb$ collisions at $\sqrt{s_{NN}} = 5.02$ TeV  \cite{cms_prel}. Such comparison is also performed here for the first time. As we will demonstrate in what follows, our results indicate that the analysis of the $t$ -- spectrum can be useful to discriminate between the different approaches to gluon saturation effects. Moreover, we will show that these models are able to describe the CMS data at small -- $t$ but  underestimate the few data points at large -- $t$.

\begin{figure}[t]
\begin{center}
\scalebox{0.45}{\includegraphics{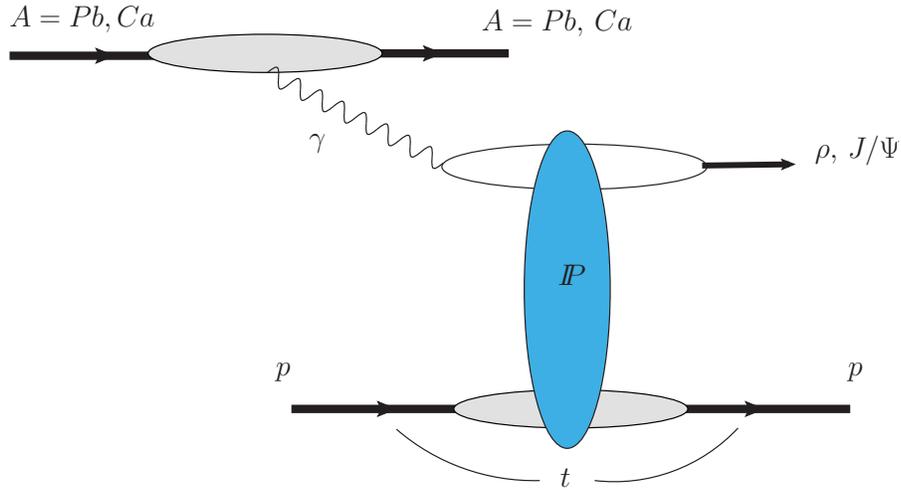}}
\caption{Exclusive vector meson photoproduction in $pA$ collisions.}
\label{fig:diagrama}
\end{center}
\end{figure}

Initially, let's present a brief review of the formalism used in our calculations. The exclusive vector meson photoproduction in $pA$ collisions is dominated by photon - proton interactions, since the nuclear photon flux is enhanced by the square of the nuclear charge ($Z$) \cite{upc}. The process is represented in Fig. \ref{fig:diagrama}.
The final state will be characterized by two intact hadrons ($A$ and $p$) and two rapidity gaps, i.e. the outgoing particles ($A$, $V = \rho, \, J/\Psi$ and $p$) are separated by a large region in rapidity in which there is no additional hadronic activity observed in the detector.  The differential cross section can be expressed as follows
\begin{eqnarray}
\frac{d\sigma \,\left[A + p \rightarrow   A \otimes V \otimes p\right]}{dY\,dt}  =  n_A(\omega) \, \cdot \, \frac{d\sigma}{dt}(\gamma p 
\rightarrow V \otimes p)\,\,\,,
\label{dsigdy}
\end{eqnarray}
where the rapidity ($Y$) of the vector meson in the final state is determined by the photon energy $\omega$ in the collider frame and by the mass $M_{V}$ 
of the vector meson [$ Y \propto \ln \, ( \omega/M_{V})$]. Moreover, $d\sigma/dt$ is the differential cross section of the $\gamma p \rightarrow V \otimes p$ process, with the symbol $\otimes$ representing the presence of a rapidity gap in the final state. Furthermore, $n_A(\omega)$ denotes the  equivalent photon 
spectrum  of the relativistic incident nucleus. As in our previous studies \cite{run2,nos_tdist} we will assume   a point -- like form factor for the nucleus, which implies that \cite{upc}
 \begin{eqnarray}
   n_{A}(\omega) = \frac{2Z^{2}\alpha_{em}}{\pi }  
\left[
\xi K_{0}(\xi) K_{1}(\xi) -\frac{\xi^{2}}{2} \left( K_{1}^{2}(\xi) - K_{0}^{2}(\xi)  
\right )  \right]  ,
  \end{eqnarray}
where $
 \xi = \omega \left( R_{A} + R_{p} \right) / \gamma_{L}$, 
with $\gamma_L$ being the Lorentz factor. The differential cross section for the 
$\gamma p 
\rightarrow V \otimes p$ process is given by
\begin{eqnarray}
\frac{d\sigma}{dt}
& = & \frac{1}{16\pi}  |{\cal{A}}^{\gamma p \rightarrow V p }(x,  \Delta)|^2\,\,,
\label{dsigdt}
\end{eqnarray}
where  ${\cal{A}}$ is the  amplitude for producing an exclusive vector meson diffractively. In the color dipole formalism \cite{nik}, this amplitude can be factorized in terms of the fluctuation of the virtual photon into a $q \bar{q}$ color dipole, the dipole-hadron scattering by a color singlet exchange (denoted $\pom$ in  Fig. \ref{fig:diagrama})  and the recombination into the vector meson  $V$. Consequently, the amplitude can be expressed as follows
\begin{eqnarray}
 {\cal A}^{\gamma p \rightarrow V p }({x},\Delta)  =  i
\int dz \, d^2\rr \, d^2\rb \,  e^{-i[\rb-(1-z)\rr].\rd}  
 \,\, (\Psi^{V*}\Psi)  \,\,2 {\cal{N}}^p({x},\rr,\rb) \,\,,
\label{amp}
\end{eqnarray}
where  $(\Psi^{V*}\Psi)$ denotes the wave function overlap between the  photon and vector meson wave functions, $\Delta = - \sqrt{t}$ is the momentum 
transfer and $\rb$ is the impact parameter of the dipole relative to the proton target. Moreover, the variables  $\rr$ and $z$ are the dipole transverse pair separation and the momentum fraction of the photon carried by a quark (an antiquark carries then $1-z$), respectively. As in Ref. \cite{nos_tdist}, in what follows we will consider the Boosted 
Gaussian model \cite{KT,KMW} for the overlap function.
The function $ {\cal N}^p (x, \rr, \rb)$ is the 
forward dipole-proton scattering amplitude (for a dipole at  impact parameter $\rb$) which encodes all the information about the hadronic scattering. It  depends on the $\gamma h$  center - of - mass reaction energy, 
$W = [2 \omega \sqrt{s_{NN}}]^{1/2}$, through the variable $ x = M^2_V/W^2$.
One of the main open questions in QCD is the treatment of its high energy regime, where non -- linear (gluon saturation) effects are expected to contribute \cite{hdqcd}. Currently, the bCGC and IP-Sat models, which are based on different assumptions for the treatment of the gluon saturation effects,  describe with success the high precision HERA data for inclusive and exclusive processes.  In  the impact parameter Color Glass Condensate (bCGC) model \cite{KMW} the dipole - proton scattering amplitude is given by 
\begin{widetext}
\begin{eqnarray}
\mathcal{N}^p(x,\rr,\rb) =   
\left\{ \begin{array}{ll} 
{\mathcal N}_0\, \left(\frac{ r \, Q_s(b)}{2}\right)^{2\left(\gamma_s + 
\frac{\ln (2/r \, Q_s(b))}{\kappa \,\lambda \,Y}\right)}  & \mbox{$r Q_s(b) \le 2$} \\
 1 - e^{-A\,\ln^2\,(B \, r \, Q_s(b))}   & \mbox{$r Q_s(b)  > 2$} \,\,,
\end{array} \right.
\label{eq:bcgc}
\end{eqnarray}
\end{widetext} 
with  $\kappa = \chi''(\gamma_s)/\chi'(\gamma_s)$, where $\chi$ is the 
LO BFKL characteristic function.  The coefficients $A$ and $B$  
are determined uniquely from the condition that $\mathcal{N}^p(x,\rr,\rb)$, and its derivative 
with respect to $r\,Q_s(b)$, are continuous at $r\,Q_s(b)=2$. The impact parameter dependence of the  proton saturation scale $Q_s(b)$  is given by:
\begin{equation} 
  Q_s(b)\equiv Q_s(x,b)=\left(\frac{x_0}{x}\right)^{\frac{\lambda}{2}}\;
\left[\exp\left(-\frac{{b}^2}{2B_{\rm CGC}}\right)\right]^{\frac{1}{2\gamma_s}},
\label{newqs}
\end{equation}
with the parameter $B_{\rm CGC}$  being obtained by a fit of the $t$-dependence of 
exclusive $J/\psi$ photoproduction. The  factors $\mathcal{N}_0$ and  $\gamma_s$  were  
taken  to be free. In what follows we consider the set of parameters obtained in 
Ref. \cite{amir} by fitting the recent HERA data on the reduced $ep$ cross sections:
 $\gamma_s = 0.6599$, $\kappa = 9.9$, $B_{CGC} = 5.5$ GeV$^{-2}$, $\mathcal{N}_0 = 0.3358$, $x_0 = 0.00105$ and $\lambda = 0.2063$. In the bCGC model, the saturation regime, where $r Q_s(b)  > 2$, is described by the Levin - Tuchin law \cite{levin_tuchin} and the linear one by the BFKL dynamics near of the saturation line.
On the other hand, in the IP-Sat model \cite{ipsat2,ipsat3}, ${\cal N}^p$
 has  an eikonalized 
form  and  depends on a gluon distribution evolved via DGLAP equation, being  given 
by 
\begin{eqnarray}
 {\cal N}^p(x,\mbox{\textbf{\textit{r}}},\mbox{\textbf{\textit{b}}}) = 
 1 - \exp \left[
\frac{\pi^{2}r^{2}}{N_{c}} \alpha_{s}(\mu^{2}) \,\,xg\left(x, \frac{4}{r^{2}} + 
\mu_{0}^{2}\right)\,\, T_{G}(b) 
 \right] ,
 \label{ipsat}
\end{eqnarray}
with a  Gaussian profile
\begin{eqnarray}
T_{G}(b) = \frac{1}{2\pi B_{G}}  
\exp\left(-\frac{b^{2}}{2B_{G}} \right) .
\end{eqnarray}
The initial gluon distribution evaluated at $\mu_{0}^{2}$ is taken to be $
xg(x,\mu_{0}^{2}) =  A_{g}x^{-\lambda_{g}} (1-x)^{5.6}$. In this work we 
assume the parameters obtained in Ref. \cite{ipsat4}.  One have that as in the bCGC model, the IP-Sat predicts the saturation of $ {\cal N}^p$ at high energies and/ot large dipoles, but the approach to this regime is not described by the Levin - Tuchin law. Moreover, in contrast to the bCGC model, the IP-Sat takes into account the effects associated to the  DGLAP evolution, which are expected to be important in the description of the small dipoles.
Consequently, both models are based on different assumptions for the linear and non - linear regimes. As pointed above, the current high precision HERA data are not able to discriminate between these models.  In what follows we analyze the possibility of constraining the models of  gluon saturation effects in  exclusive vector meson photoproduction at $pA$ collisions.

\begin{figure}[t]
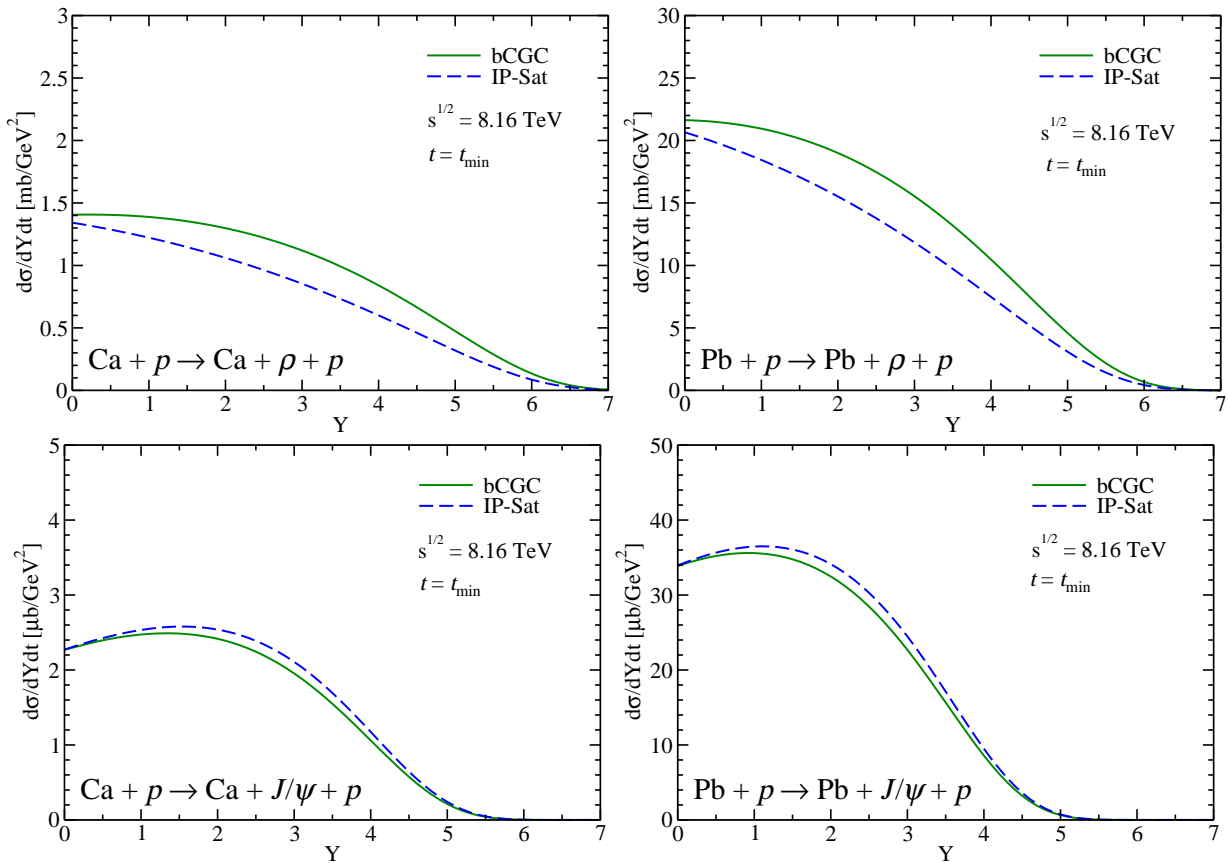

\begin{center}
\scalebox{0.33}{\includegraphics{dsdtdY_pCa_boosted_y_dist_8160_linear.eps}}
\scalebox{0.33}{\includegraphics{dsdtdY_pPb_boosted_y_dist_8160_linear.eps}} \\
\scalebox{0.33}{\includegraphics{dsdtdY_pCa_boosted_jpsi_y_dist_8160_linear.eps}}
\scalebox{0.33}{\includegraphics{dsdtdY_pPb_boosted_jpsi_y_dist_8160_linear.eps}}

\caption{Rapidity distribution for the exclusive $\rho$ (upper panels)  and $J/\Psi$ (lower panels) photoproduction in $pCa$ and $pPb$ collisions at $\sqrt{s_{NN}} = 8.16$ TeV.}
\label{Fig:rap_pA}
\end{center}
\end{figure}

Let's consider the  exclusive $\rho$ and $J/\Psi$  photoproduction in $pCa$ and $pPb$ collisions at the LHC energies. Our main focus will be on the transverse momentum distributions, which are expected to be studied considering the higher statistics of  Run 2 and 3 
\cite{review_forward}. However, firstly let us analyse the impact of the gluon saturation effects on the rapidity distributions at a fixed value of the momentum transfer $t$. We will estimate  Eq. (\ref{dsigdy}) for $t = t_{min}$, with $t_{min} = - m_N^2 M_V^4/W^4$. In Fig. \ref{Fig:rap_pA} we present our predictions for the rapidity distributions considering the exclusive $\rho$ (upper panels) and $J/\Psi$ (lower panels) photoproduction  in 
$pCa$  and $pPb$ collisions. We observe that the difference between the bCGC and IP-Sat is larger  
for $\rho$ production, with the IP-Sat predictions being smaller than the bCGC ones. On the other hand, the IP-Sat model predicts larger values of the rapidity distribution when the $J/\Psi$ production is considered. 
These results are expected, since the bCGC and IP-Sat  models assume different behavior for the linear and non - linear regimes. In the $\rho$ case, the process is dominated by the contribution of large dipole sizes, which are expected to be strongly suppressed by the gluon saturation effects. On the other hand,  $J/\Psi$ production is dominated by small dipoles, i. e. the cross section is expected to be mainly determined by the linear regime of the QCD dynamics. The main difference between the predictions for $pCa$ and $pPb$ collisions is the normalization of the distributions. This result is also expected, since  the distribution is calculated by the product of the photon flux and the photon - proton cross section [See Eq. (\ref{dsigdy})], with $n_A$ being proportional to $Z^2$. The rapidity and transverse momentum dependencies are determined by the  $\gamma p \rightarrow V p$ cross section, which is the same for $pCa$ and $pPb$ collisions. Consequently, in what follows, we will only  present our predictions for the $t$-- distributions in $pPb$ collisions.

\begin{figure}[t]
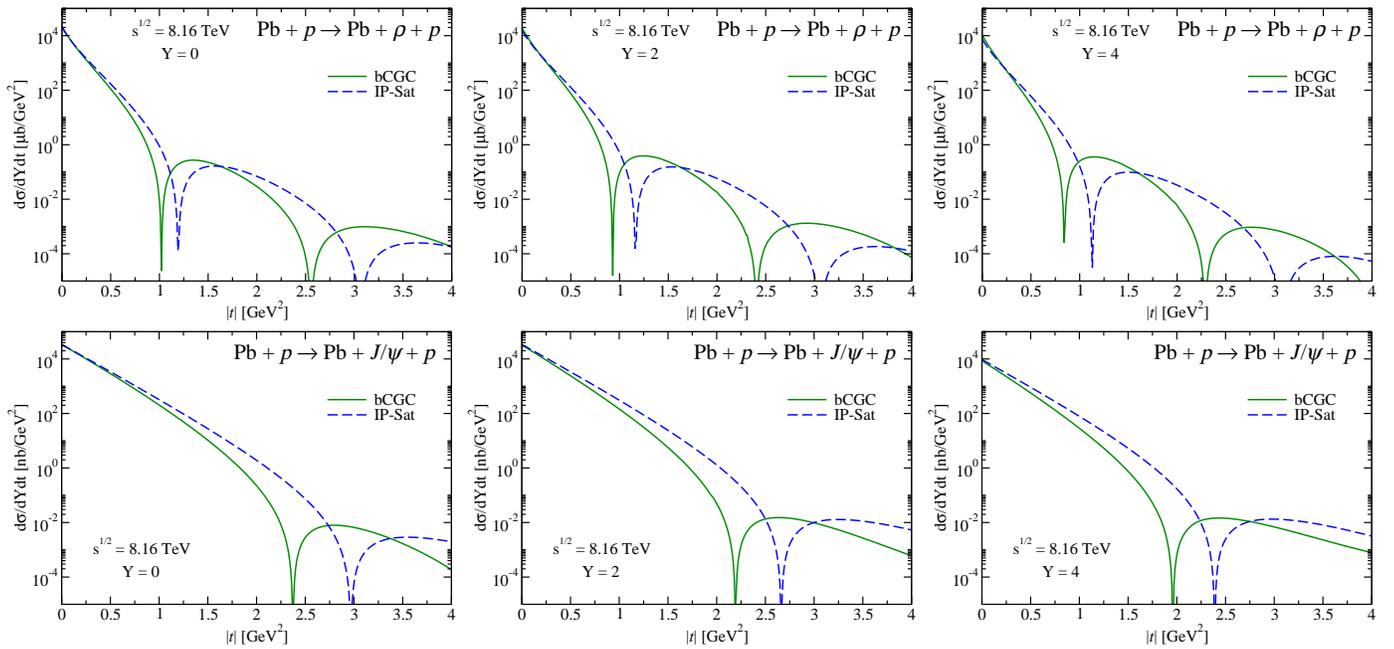

\begin{center}
\begin{tabular}{ccc}
\scalebox{0.24}{\includegraphics{dsdtdY_pPb_boosted_t_dist_8160_y_0.eps}} &
\scalebox{0.24}{\includegraphics{dsdtdY_pPb_boosted_t_dist_8160_y_2.eps}} &
\scalebox{0.24}{\includegraphics{dsdtdY_pPb_boosted_t_dist_8160_y_4.eps}} \\
\scalebox{0.24}{\includegraphics{dsdtdY_pPb_boosted_jpsi_t_dist_8160_y_0.eps}} &
\scalebox{0.24}{\includegraphics{dsdtdY_pPb_boosted_jpsi_t_dist_8160_y_2.eps}} &
\scalebox{0.24}{\includegraphics{dsdtdY_pPb_boosted_jpsi_t_dist_8160_y_4.eps}}
\end{tabular}
\caption{Transverse momentum distributions for the exclusive $\rho$ (upper panels) and $J/\Psi$ (lower panels) photoproduction in $pPb$ collisions at $\sqrt{s_{NN}} = 8.16$ TeV assuming three different values for the  vector meson rapidity.}
\label{Fig:dsigdt_tdist}
\end{center}
\end{figure}

Let us now  analyze the predictions of the  different gluon saturation models for the transverse momentum distributions considering $pPb$ collisions at $\sqrt{s_{NN}} = 8.16$ TeV and assuming three different fixed values for the vector meson rapidity ($Y = 0$, 2 and 4).  Our results for the exclusive $\rho$ and $J/\Psi$ photoproduction are presented in the upper and lower panels of  Fig. \ref{Fig:dsigdt_tdist}, respectively. We observe that the bCGC and IP-Sat predictions are similar at small - $|t|$, but differ at larger values. The position of the dip is dependent on the description of the gluon saturation effects, with the bCGC model predicting the dip at smaller values of $|t|$, independently of the produced vector meson. Moreover, we see that the position of the dip is displaced at smaller $|t|$ with the growth of the rapidity and the number of dips predicted for the $\rho$ production in the range $|t| \le 4$ GeV$^2$ is larger than for the $J/\Psi$ case. These results indicate that the study of the $t$ - distribution in the range  $0.75 \le |t| \le 1.5$ GeV$^2$  ($2.0 \le |t| \le 3.0$ GeV$^2$) for the case of $\rho$ ($J/\Psi$) production can be useful to contrain the description of the gluon saturation effects.

\begin{figure}[t]
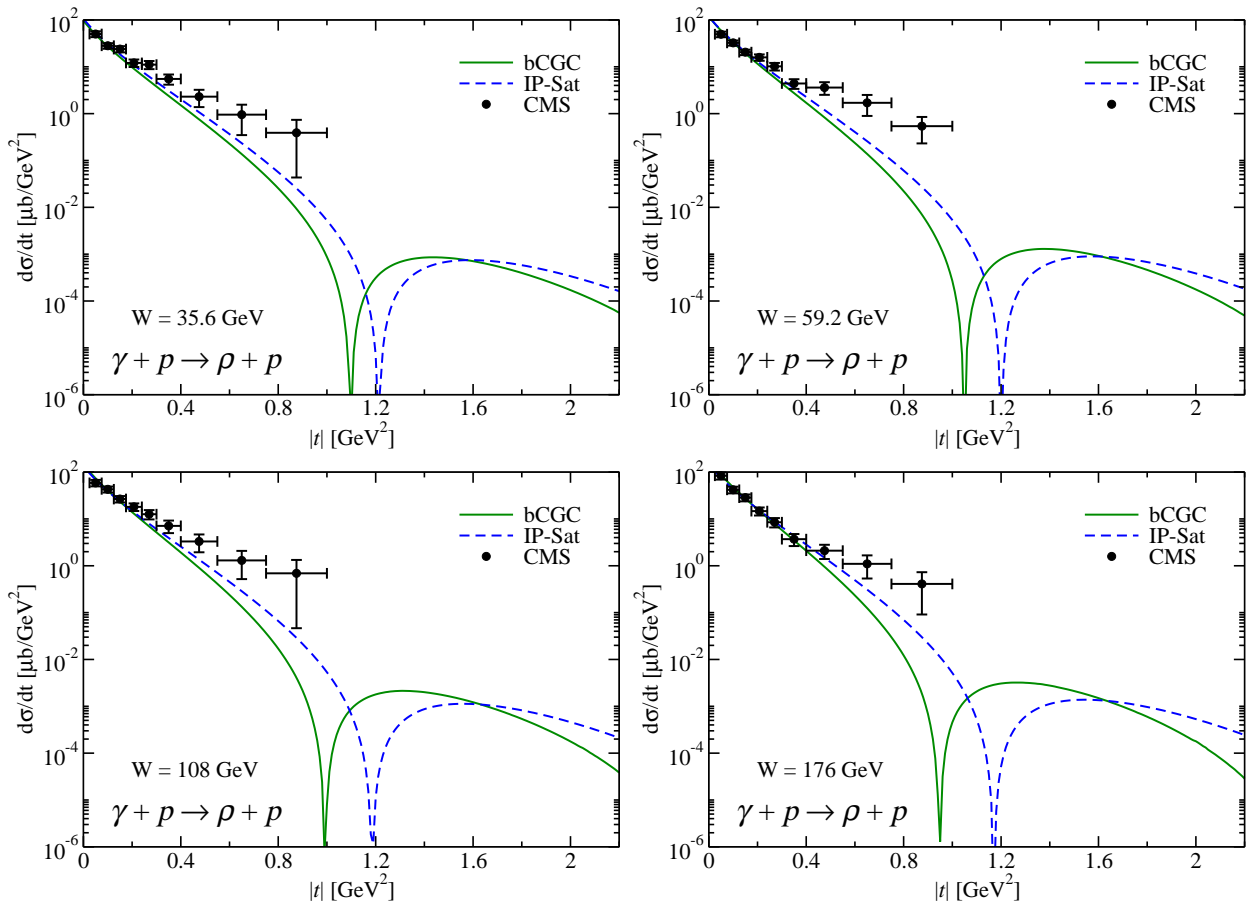

\begin{tabular}{cc}
  \includegraphics[scale=0.33]{dsdt_rho_gp_W36_boosted.eps} &
 \includegraphics[scale=0.33]{dsdt_rho_gp_W59_boosted.eps} \\
  \includegraphics[scale=0.33]{dsdt_rho_gp_W108_boosted.eps} &
 \includegraphics[scale=0.33]{dsdt_rho_gp_W176_boosted.eps}
 \end{tabular}
\caption{Transverse momentum distributions for the exclusive $\rho$ photoproduction at different  center - of - mass energies of the $\gamma p $ system. Preliminary data from the CMS Collaboration \cite{cms_prel}.}
\label{Fig:dsigdt_ep}
\end{figure}

 The exclusive $\rho$ photoproduction in ultraperipheral $pPb$ collisions at $\sqrt{s_{NN}} = 5.02$ TeV has been studied by the CMS Collaboration. In particular, they released, very recently, the first (preliminary) data \cite{cms_prel} for the $t$ -- distributions of the $\gamma p \rightarrow \rho p$ process at different center -- of -- mass energies of the $\gamma p$ system. Assuming that the nuclear photon flux is well known and that there is a direct relation between the rapidity $Y$ of the vector meson and the $\gamma p$ center -  of - mass energy ($W$), they measured $d\sigma/dt$ for different rapidity bins and, consequently, for averaged values of $W$.
 A comparison between our predictions and these preliminary data is presented in Fig. \ref{Fig:dsigdt_ep}. The bCGC and IP-Sat models describe quite well  the distributions for $|t| \le 0.4$ GeV$^2$. On the other hand, at larger values of $|t|$, where the number of experimenal points is smaller and the uncertainty is larger, the predictions of the gluon saturation models underestimate the data, with the IP-Sat predictions being closer to the data.  It is important to emphasize that the discrepancy starts to occur exactly in the region where the presence of dips becomes important and the $t$ - distribution can no longer be described by an exponential with a fixed slope.
If confirmed, these data can be a first indication that the model of the spatial distribution of  gluons in the proton (present in the bCGC and IP-Sat models) should be improved in the study of  gluon saturation effects. Certainly, more data on exclusive vector meson photoproduction will be very useful to improve our understanding of the QCD dynamics at high energies.

Finally, let's summarize our main results and conclusions. In this letter we have investigated the exclusive $\rho$ and $J/\Psi$ photoproduction in $pA$ collisions at the LHC motivated by the expectation that this process  may allow us to  constrain  the description of the QCD dynamics at high energies.
Differently from $pp$ and $AA$ collisions, in $pA$ collisions the rapidity of the vector meson allows to unambiguously determine the $\gamma p$ center - of - mass energy and, consequently, to probe the QCD dynamics at the given value of the Bjorken -- $x$ variable. We have considered  $\rho$ and $J/\Psi$ production, which mainly probes the non - linear and linear QCD regimes, respectively, and presented the bCGC and IP-Sat predictions for the rapidity and transverse momentum distributions. 
These two models, even though describing the available HERA data, are based on different assumptions for the gluon saturation effects. We demonstrated that their predictions for the $t$ -- spectra are similar at small values of $|t|$ but differ at large - $|t|$, with the position of the dip being model dependent. A comparison of our predictions with the very recent (preliminary) CMS data has been presented for the first time, with the data at small - $|t|$ being quite well described by both gluon saturation models. However, the  large - $|t|$ data are underestimated by the models. This can be a first indication that the description of the spatial distribution of the gluons in the proton should be improved. 
These results indicate that the experimental analysis of the transverse momentum distribution is  useful to discriminate between different approaches for the QCD dynamics as well to improve our description of the gluon saturation effects.

%\vspace{-0.6cm}

\begin{acknowledgements}
 VPG acknowledges useful discussions with  J. Cepila, J. G. Contreras,  J. D. Tapia - Takaki and W. Schafer.  VPG is grateful to the members of the Department of Physics and Astronomy of the University of Kansas by the warm hospitality during the initial phase of this study.
 This work was  partially financed by the Brazilian funding
agencies CNPq,  FAPERGS, FAPESP and INCT-FNA (process number 
464898/2014-5).

\end{acknowledgements}

\end{document}